\documentclass[aps,prl,twocolumn,nofootinbib,tightenlines,showpacs,%
floatfix]{revtex4}

\usepackage{epsfig,bm}

\begin{document}

\preprint{hep-ph/0310117}


\title{Properties of Baryon Anti-decuplet}


\author{Yongseok Oh}%
\email{yoh@phya.yonsei.ac.kr}

\author{Hungchong Kim}%
\email{hung@phya.yonsei.ac.kr}

\author{Su Houng Lee}%
\email{suhoung@phya.yonsei.ac.kr}

\affiliation{Institute of Physics and Applied Physics,
Yonsei University, Seoul 120-749, Korea}


\begin{abstract}
We study the group structure of baryon anti-decuplet containing
the $\Theta^+$. We derive the SU(3) mass relations among the
pentaquark baryons in the anti-decuplet, when there is either no
mixing or ideal mixing with the pentaquark octet, as advocated by
Jaffe and Wilczek. This constitutes the Gell-Mann--Okubo mass
formula for the pentaquark baryons. We also derive SU(3) symmetric
Lagrangian for the interactions of the baryons in the
anti-decuplet with the meson octet and the baryon octet. Our
analysis for the decay widths of the anti-decuplet states suggests
that the $N(1710)$ is ruled out as a pure anti-decuplet state, but
it may have anti-decuplet component in its wavefunction if the
multiplet is mixed with the pentaquark octet.
\end{abstract}


\maketitle


The recent discovery of the $\Theta^+$ baryon by LEPS Collaboration at
SPring-8 \cite{LEPS03}, which has been confirmed by several groups
\cite{CLAS03-b-SAPHIR03-DIANA03-ADK03}, initiated a lot of theoretical
works in the field of exotic hadrons.
Experimentally, the $\Theta^+$ is observed to have a mass of 1540 MeV and
a decay width of $< 25$ MeV.
Because of its positive strangeness, the $\Theta^+$ baryon is exotic since
its minimal quark content should be $uudd\bar{s}$.
Other states that have positive strangeness but different charges are not
observed, which suggests that the $\Theta^+$ is an isosinglet.
The existence of such an exotic state with narrow width was first predicted
by Diakonov {\it et al.\/} \cite{DPP97} in the chiral soliton model,
where the $\Theta^+$ is a member of the baryon anti-decuplet.
Although it should be confirmed by other experiments, the recently
discovered $\Xi_{3/2}^{--}$ baryon \cite{NA49-03} strongly
supports the anti-decuplet nature of pentaquark baryons.
The discussion on the existence of a baryon anti-decuplet has a longer
history going back to the early 1970's \cite{Golo71}, and
in the Skyrme model \cite{Chem85}.
Pentaquark states with a heavy antiquark ($uudd\bar{c}$, $uudd\bar{b}$)
were also predicted in the Skyrme model.
Here an important issue is whether such nonstrange heavy pentaquark states
are stable against strong decays
\cite{RS93,OPM94b-OPM94c,OP95,GSR87-Lip87,JW03}.
Subsequent theoretical investigations on the $\Theta^+$ include approaches
based on the constituent quark model
\cite{SR03,CCKN03,Cheung03-Gloz03a-KL03a},
Skyrme model \cite{Weig98-WK03,Pras03,JM03,BFK03,IKOR03}, QCD sum rules
\cite{MNNRL03,Zhu03-SDO03}, lattice QCD \cite{CFKK03-Sasaki03},
chiral potential model \cite{Hosaka03},
large $N_c$ QCD \cite{Cohen03-CL03},
and Group theory approach \cite{Wyb03}.
The production of the $\Theta^+$ was also discussed in relativistic nuclear
collisions \cite{Rand03-CGKLL03}, where the number of the
anti-$\Theta^+(1540)$ produced are expected to be similar to that of
the $\Theta^+(1540)$.

Several pressing issues that should be clarified are whether the
$\Theta^+$ has positive or negative parity,
and whether the Roper $N(1710)$ is included in the
baryon anti-decuplet with the $\Theta^+$.
As for the spin-parity of the $\Theta^+$, several works claim that
$J^P=\frac12^-$ is more natural \cite{CCKN03,Zhu03-SDO03,CFKK03-Sasaki03},
which, however, is in contrast to the prediction of the soliton models,
where the relative orbital angular momentum plays an important
role \cite{OP95,JW03}.
The $J^P=\frac12^+$ assignment is also consistent with the recent
study of the quark model \cite{CCKN03b}, which predicts the lowest mass
pentaquark state to be in the $P$ wave state.
The elementary production processes of the $\Theta^+$ have been studied
to address this question \cite{LK03a-LK03b-NHK03-OKL03a}.
As for the structure of the anti-decuplet states, Jaffe and Wilczek
proposed that the pentaquark is a diquark-diquark-antiquark state which
is ideally mixed with the corresponding pentaquark in the octet \cite{JW03}.
In this approach, the $N(1710)$ is classified as a member of the
pentaquark states as in the chiral soliton model of Ref. \cite{DPP97},
but the mass spectrum is totally different, as the mixing introduces the
$N(1440)$ as the lowest state among the pentaquark states.
However, it is also claimed that such interpretation is doubted as
the Roper resonances fall into the (excited) ${\bf 56}$-plets of
SU(6) \cite{Gloz03b}.

In this paper, we study the SU(3) flavor structure of the baryon
anti-decuplet. We derive the SU(3) mass relations among the
baryons in the anti-decuplet when there is either no mixing or
ideal mixing with the pentaquark octet, where the later mixing
scenario was advocated by Jaffe and Wilczek \cite{JW03}.   This
constitutes the Gell-Mann--Okubo mass formula for the pentaquark
baryons.  We also derive the SU(3) symmetric interactions of the
anti-decuplet with the baryon octet and pseudoscalar meson octet.

\begin{figure}
\centering
\epsfig{file=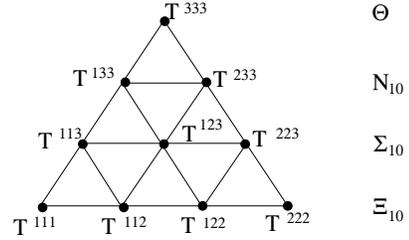, width=0.6\hsize}
\caption{A baryon anti-decuplet.}
\label{diag}
\end{figure}

As shown in Fig.~\ref{diag}, we start by introducing the irreducible
tensor notation $T^{jkl}$ to represent the baryons in the anti-decuplet,
which are denoted as the (0,3) representation in the usual $(p,q)$
notation \cite{Low}.
Note that the baryon decuplet is in the (3,0) and represented by $T_{jkl}$.
Since $T^{jkl}$ is an eigenstate of the hypercharge $Y$ and the third
component of the isospin $I_3$, one can directly match it with the
physical baryon states. (See Table~\ref{table1}.)

\begin{table}
\centering
\begin{ruledtabular}
\begin{tabular}{c|ccc|c}
   & $Y$ & $I_3$ & $Q$ & Particle \\ \hline
$T^{111}$ & $-1$ & $-\frac32$ & $-2$ & $\sqrt6 \Xi^{--}_{10}$ \\
$T^{112}$ & $-1$ & $-\frac12$ & $-1$ & $-\sqrt2 \Xi^{-}_{10}$ \\
$T^{113}$ & $0$ & $-1$ & $-1$ & $\sqrt2 \Sigma_{10}^{-}$ \\
$T^{122}$ & $-1$ & $\frac12$ & $0$ & $\sqrt2 \Xi^{0}_{10}$ \\
$T^{123}$ & $0$ & $0$ & $0$ & $-\Sigma_{10}^{0}$ \\
$T^{133}$ & $1$ & $-\frac12$ & $0$ & $\sqrt2 N_{10}^{0}$ \\
$T^{222}$ & $-1$ & $\frac32$ & $1$ & $-\sqrt6 \Xi^{+}_{10}$ \\
$T^{223}$ & $0$ & $1$ & $1$ & $-\sqrt2 \Sigma_{10}^{+}$ \\
$T^{233}$ & $1$ & $\frac12$ & $1$ & $-\sqrt2 N_{10}^{+}$ \\
$T^{333}$ & $2$ & $0$ & $1$ & $\sqrt6 \Theta^{+}$ \\
\end{tabular}
\end{ruledtabular}
\caption{Hypercharge, third component of isospin, charge, and physical
state with proper normalization of the baryon anti-decuplet members. The
phases are chosen to be consistent with Ref.~\cite{deS63}.}
\label{table1}
\end{table}

With these informations, we study baryon masses using SU(3) flavor
symmetry.
The well-known Gell-Mann--Okubo mass formula reads
\begin{equation}
M = M_0 + \alpha Y + \beta D^3_3,
\label{mass-form}
\end{equation}
where $M_0$ is a common mass of a given multiplet and $D^3_3 =
I(I+1) - Y^2/4 - C/6$, with $C = 2(p+q) + \frac23 (p^2 + pq +
q^2)$ for the $(p,q)$ representation.
$\alpha$ and $\beta$ are mass constants that are in principle different
for different multiplets.
Using these constants, one can write down the masses of all the baryon
within a multiplet.
Moreover, using these mass formulas, one can derive the well-known
Gell-Mann--Okubo mass relation, the decuplet equal spacing rule, and the
hyperfine splitting rule,
\begin{eqnarray}
&& 3 M_\Lambda + M_\Sigma = 2(M_N + M_\Xi), \label{rel:GMO}
\\
&& M_\Omega - M_{\Xi^*} = M_{\Xi^*} - M_{\Sigma^*} = M_{\Sigma^*}
- M_\Delta, \label{rel:DES}
\\
&& M_{\Sigma^*} - M_\Sigma + \frac32 (M_\Sigma - M_\Lambda) =
M_\Delta - M_N. \label{rel:HYP}
\end{eqnarray}
The last relation follows only if we assume $\alpha_8=\alpha_{10}$
and $\beta_8=\beta_{10}$, where the subscripts $8,10$ represent the
octet and decuplet representation respectively.

Using Eq.~(\ref{mass-form}), the mass formulas for the baryon
anti-decuplet can also be written down as follows,
\begin{eqnarray}
M_\Theta &=& M_{\overline{10}} + 2 (\alpha_{\overline{10}} -
\frac32 \beta_{\overline{10}}), \nonumber \\
M_{N_{10}} &=& M_{\overline{10}} + (\alpha_{\overline{10}} -
\frac32 \beta_{\overline{10}}), \nonumber \\
M_{\Sigma_{10}} &=& M_{\overline{10}}, \nonumber \\
M_{\Xi_{10}} &=& M_{\overline{10}} - ( \alpha_{\overline{10}} -
\frac32 \beta_{\overline{10}}). \label{mass:10bar}
\end{eqnarray}

We now discuss the anti-decuplet mass spectrum.

{\bf Pure anti-decuplet}:  Eq. (\ref{mass:10bar}) shows that the
equal spacing rule holds also for the baryon anti-decuplet,
\begin{equation}
M_{\Xi_{10}} - M_{\Sigma_{10}} = M_{\Sigma_{10}} - M_{N_{10}} =
M_{N_{10}} - M_\Theta, \label{rel:ADES}
\end{equation}
which was already used in Ref. \cite{DPP97}.
This suggests that if any two of the masses in the anti-decuplet are known,
the other masses can be predicted by the mass relation.
Using $M_\Theta= 1540$ MeV and identifying the recent measurement of
$M_{\Xi_{3/2}} = 1862$ MeV by the NA49 experiment \cite{NA49-03} as
$M_{\Xi_{10}}$, we have
\begin{eqnarray}
&& M_\Theta = \underline{1540} \mbox{ MeV}, \quad M_{N_{10}} =
1647 \mbox{ MeV},
\nonumber \\
&& M_{\Sigma_{10}} = 1755 \mbox{ MeV}, \quad M_{\Xi_{10}} =
\underline{1862}  \mbox{ MeV}, \label{mass-10}
\end{eqnarray}
where the masses used as input are underlined.

It should be noted that for the anti-decuplet, there are in fact
only two independent parameters;  $M_{\overline{10}}$ and $\delta
m_{\overline{10}}= (\alpha_{\overline{10}} - \frac32
\beta_{\overline{10}})$.
This is also true for the decuplet;
$M_{10}$ and $\delta m_{10}= (\alpha_{10} + \frac32 \beta_{10})$.
However, the mass parameters are quite different for the
anti-decuplet $\delta m_{\overline{10}} \sim - 107 $ MeV, and the
decuplet $\delta m_{10} \sim -150 $ MeV.
Moreover, if we assume $\alpha_{\overline{10}}=\alpha_{10}=\alpha_8$ and
$\beta_{\overline{10}}=\beta_{10}=\beta_8$, we would get an
unrealistically large mass splitting $\delta m_{\overline{10}}
\sim -240$ MeV.
The problem comes from using the parameters obtained from baryons with
valence strange quarks only in the mass splitting for pentaquark baryon,
where the anti-strange quark is also important in giving their mass
splitting.
That is why Jaffe and Wilczek introduced a phenomenological mass formula for
the pentaquark states where the important mass splitting comes from
not only terms proportional to the number of strange quarks $n_s$
but also from terms proportional to $n_{\bar{s}}$ \cite{JW03},
\begin{equation}
\delta M = \gamma (n_s + n_{\bar s}),
\label{phen-mass}
\end{equation}
which, however, does not change the mass relations
(\ref{rel:GMO})-(\ref{rel:HYP}).

{\bf Ideal mixing with pentaquark octet}: Let us consider the
mixing of the anti-decuplet with the pentaquark octet \cite{JW03}.
Using Eq. (\ref{mass-form}), we can write the masses of the
pentaquark octet as
\begin{eqnarray}
M_{N_8} &=& M_8' + \alpha_8' - \frac12 \beta_8', \nonumber \\
M_{\Sigma_8} &=& M_8' + \beta_8', \nonumber \\
M_{\Xi_8} &=& M_8' - \alpha_8' - \frac12 \beta_8', \nonumber \\
M_{\Lambda_8} &=& M_8' - \beta_8'.
\label{mass:8'}
\end{eqnarray}

As in the model of Jaffe and Wilczek \cite{JW03}, we assume that
the pentaquark octet and anti-decuplet are ideally mixed so that
the $s\bar{s}$ component is isolated,
\begin{eqnarray}
&& N_{10} = \sqrt{\frac{1}{3}} N_q+ \sqrt{\frac{2}{3}} N_s,
\quad
N_8  =  \sqrt{\frac{2}{3}} N_q- \sqrt{\frac{1}{3}} N_s ,  \nonumber \\
&& \Sigma_{10} = \sqrt{\frac{2}{3}} \Sigma_q + \sqrt{\frac{1}{3}}
\Sigma_s, \quad
\Sigma_8 = \sqrt{\frac{1}{3}} \Sigma_q- \sqrt{\frac{2}{3}}
\Sigma_s,
\nonumber \\
\end{eqnarray}
where the subscript $q$ ($s$) means that the pentaquark has only
light (strange) antiquark.
Based on the above mixing formula, one can show that the original mass
term in the Hamiltonian, if it has mixing between octet and
anti-decuplet ($c_1$ and $c_2$ terms), is diagonalized as
\begin{eqnarray}
H_{m} & =&  M_{N_{10}} {N_{10}}^2 + M_{N_8}
{N_8}^2 + c_1 N_{10} N_8 \nonumber \\
&& +  M_{\Sigma_{10}} {\Sigma_{10}}^2 + M_{\Sigma_8}
{\Sigma_8}^2 + c_2 {\Sigma_{10}} \Sigma_8 \nonumber \\
&=& (2 M_{N_8} - M_{N_{10}} ) {N_q}^2 + (2 M_{N_{10}} - M_{N_8})
{N_s}^2 \nonumber \\
&&+(2 M_{\Sigma_{10}} - M_{\Sigma_8} ) {\Sigma_q}^2 + (2
M_{\Sigma_8} - M_{\Sigma_{10}}) {\Sigma_s}^2.
\nonumber \\
\end{eqnarray}
Then we have
\begin{eqnarray}
M_{N_q} &=& 2( M_8'+ \alpha_8' -\frac12 \beta_8') - (M_{\overline{10}} + \delta m_{\overline{10}}), \nonumber \\
M_\Theta &=& M_{\overline{10}} + 2 \delta m_{\overline{10}}, \nonumber \\
M_{\Lambda_8} &=& M_{8}' - \beta_8', \nonumber \\
M_{\Sigma_q} &=& 2M_{\overline{10}} - (M_8' + \beta_8') , \nonumber \\
M_{N_s} &=& 2 (M_{\overline{10}}+\delta m_{\overline{10}}) - (M_8'
+ \alpha_8' - \frac12 \beta_8'),
\nonumber \\
M_{\Xi_{10}} &=& M_{\overline{10}} -\delta m_{\overline{10}}
\nonumber \\
M_{\Xi_8} &=& M_8' - \alpha_8' - \frac12 \beta_8' , \nonumber \\
M_{\Sigma_s} &=& 2( M_8'+\beta_8')  - M_{\overline{10}}.
\label{mass:physical}
\end{eqnarray}

There are eight masses and five parameters, hence three SU(3) mass
relations.
These are the generalized Gell-Mann--Okubo mass formula and
anti-decuplet equal spacing rule and read
\begin{eqnarray}
&& M_{N_q} + 2 M_{N_s} = 2 M_\Theta + M_{\Xi_{10}},
\label{rel:GMOd}
\\
&& 2 M_{\Sigma_q} + M_{\Sigma_s} = M_\Theta+ 2 M_{\Xi_{10}},
\label{rel:DESad}
\\
&& 3 M_{\Lambda_8}= M_{\Sigma_q} + M_{N_q} + 2 M_{\Xi_8} -
M_{\Xi_{10}}. \label{rel:HYPad}
\end{eqnarray}
These are the model independent SU(3) mass relations that should
be satisfied by any model calculations.

So far, only two masses are known. To go further, we need some
assumptions. First of all we will assume $M_{\Xi_8} =
M_{\Xi_{10}}$. Next, as in Ref. \cite{JW03}, we assume $M_{N_q}=
1440 $ MeV, namely identify it with the Roper. Then we get
$M_{N_s}=1751 $ MeV, which is close to the $N(1710)$. If we
further assume $M_{\Lambda_8}=M_{\Sigma_q}$ \cite{JW03}, we obtain
$M_{\Lambda_8}=1651 $ MeV and $M_{\Sigma_s}=1962 $ MeV. This is an
improved mass spectrum based on the assumptions of Ref.
\cite{JW03}.

Alternatively, we may turn on the phenomenological mass term in
eq. (\ref{phen-mass}) and  assume that $\alpha$, $\beta$, and
$\gamma$ are the same for all  the multiplet.  Then using $M_N$,
$M_\Sigma$, $M_\Lambda$, $M_\Theta$, and $M_{\Xi_{10}}$ to fit
these parameters, we obtain,
\begin{eqnarray}
&& M_{N_q} = 1584 \mbox{ MeV}, \quad
M_\Theta = \underline{1540} \mbox{ MeV}, \nonumber \\
&& M_{\Lambda_8} = 1647 \mbox{ MeV}, \quad
M_{\Sigma_q} = 1495 \mbox{ MeV}, \nonumber \\
&& M_{N_s} = 1679 \mbox{ MeV}, \quad
M_{\Xi_{10}} = M_{\Xi_8} = \underline{1862} \mbox{ MeV}, \nonumber \\
&& M_{\Sigma_s} =  2274 \mbox{ MeV},
\label{mass:numbers}
\end{eqnarray}
where the input masses are underlined.
This is very different from the previous results and shows the
sensitivity of the pentaquark mass spectrum on the assumptions made.

We now consider the SU(3) symmetric interaction of the
anti-decuplet ($\overline{D}$) with the baryon octet ($B$) and the
pseudoscalar meson octet ($P$).
The SU(3) symmetric interaction can be written as
\begin{equation}
\mathcal{L}_{\overline{D}PB} = -ig \bar{T}^{jkl} \gamma_5
P^j_m B^k_n \epsilon^{lmn} + \mbox{h.c.},
\end{equation}
where $P^j_m$ is the pseudoscalar meson octet and $B^k_n$ the baryon octet
of which the expressions can be found, e.g, in Refs. \cite{Low,Close,FR}.
Here we have used pseudoscalar coupling for the interaction, which can be
readily replaced by pseudovector coupling by imposing chiral symmetry in
the anti-decuplet sector.
The universal coupling constant $g$ can be determined from the $\Theta^+$
decay width as $g^2 = \Gamma_\Theta / (6.19 \mbox{ MeV})$, which
gives $g \approx 0.9$ if $\Gamma_\Theta = 5$ MeV \cite{Thetadecay}.
But we notice the possibility that $\Gamma_\Theta$ is much smaller
\cite{Thetadecay}.
In Table~\ref{table2}, we give the full $\overline{D}PB$ coupling constants,
some of which were derived in Ref. \cite{CCKN03b}.
We also note that the $\Theta^+ K^0 p$ and $\Theta^+ K^+ n$ interactions have
different phase, which differs from Ref.~\cite{DPP97},
where the two couplings have the same phase.

The next step is to find the interactions of anti-decuplet with other
multiplets.
One can obtain the anti-decuplet interaction with the meson octet as
\begin{equation}
\mathcal{L}_{\overline{D}\overline{D}P} =
g' \bar{T}^{jkl} P^j_m T^{mkl},
\end{equation}
where we have dropped the Lorentz structure of the interaction.
However, the interaction of $\overline{D}PD$, where $D$ represents
the baryon decuplet, is not allowed by SU(3) flavor symmetry.
This is because ${\bf 8} \otimes {\bf 10} = {\bf 35} \oplus {\bf 27}
\oplus {\bf 10} \oplus {\bf 8}$.
So it cannot form an $\overline{\bf 10}$ and the $\overline{D}PD$
interaction cannot be SU(3) singlet.
This gives a very strong constraint on the properties of the $N_{10}$
($\Theta^+$), since it is now prohibited to couple with $\Delta\pi$
($\Delta K$) channel.

\begin{table}
\centering
\begin{ruledtabular}
\begin{tabular}{cc|cc|cc|cc}
\multicolumn{2}{c}{$\Theta^+$} &
\multicolumn{2}{c}{$N_{10}^+$} &
\multicolumn{2}{c}{$N_{10}^0$} &
\multicolumn{2}{c}{$\Sigma_{10}^+$}
\\ \hline
$K^+ n$ & $\sqrt{6}$ & $\pi^+ n$ & $-\sqrt2$ & $\pi^0 n$ & $1$ &
$\pi^+ \Lambda$ & $-\sqrt{3}$ \\
$K^0 p$ & $-\sqrt6$ & $\pi^0 p$ & $-1$ & $\pi^- p$ & $-\sqrt2$
& $\pi^+ \Sigma^0$ & $1$ \\
  &  & $\eta p$ & $\sqrt{3}$ & $\eta n$ & $\sqrt{3}$ &
$\pi^0 \Sigma^+$ & $-1$ \\
 & & $K^+ \Lambda$ & $-\sqrt{3}$ & $K^+ \Sigma^-$ & $\sqrt2$ & $\eta
\Sigma^+$ & $\sqrt{3}$ \\
 & & $K^+ \Sigma^0$ & $1$ & $K^0 \Lambda$ &
$-\sqrt{3}$ & $K^+ \Xi^0$ & $\sqrt2$ \\
 & & $K^0 \Sigma^+$ & $\sqrt2$ & $K^0 \Sigma^0$ & $-1$ &
$\bar{K}^0 p $ & $-\sqrt2$ \\ \hline
\multicolumn{2}{c}{$\Sigma_{10}^0$} &
\multicolumn{2}{c}{$\Sigma_{10}^-$} &
\multicolumn{2}{c}{$\Xi^+_{10}$} &
\multicolumn{2}{c}{$\Xi^0_{10}$}  \\ \hline
$\pi^+ \Sigma^-$ & $-1$ & $\pi^0 \Sigma^-$ &
$1$ & $\pi^+\Xi^0$ & $\sqrt6$ & $\pi^+ \Xi^-$ & $\sqrt2$ \\
$\pi^0 \Lambda$ & $-\sqrt{3}$ & $\pi^- \Lambda$ & $-\sqrt{3}$
& $\bar{K}^0 \Sigma^+$ & $-\sqrt6$ & $\pi^0 \Xi^0$ & $-2$ \\
$\pi^- \Sigma^+$ & $1$ & $\pi^- \Sigma^0$ &
$-1$ &  &  & $\bar{K}^0 \Sigma^0$ & $2$ \\
$\eta\Sigma^0$ & $\sqrt{3}$ & $\eta \Sigma^-$ & $\sqrt{3}$ &
&  & $K^-\Sigma^+$ & $\sqrt2$ \\
$K^+ \Xi^-$ & $-1$ & $K^0 \Xi^-$ & $-\sqrt2$ & & & &  \\
${K}^0 \Xi^0$ & $-1$ & $K^- n$ & $-\sqrt2$ & & & &  \\
$\bar{K}^0 n$ & $1$ &  &  & & & &  \\
${K}^- p$ & $-1$ &  &  & & & &  \\ \hline
\multicolumn{2}{c}{$\Xi^-_{10}$} &
\multicolumn{2}{c}{$\Xi^{--}_{10}$} &
\multicolumn{2}{c}{} &
\multicolumn{2}{c}{} \\ \hline
$\pi^0 \Xi^-$ & $-2$ & $\pi^- \Xi^-$ & $-\sqrt6$ & & & & \\
$\pi^- \Xi^0$ & $-\sqrt2$ & $K^- \Sigma^-$ & $-\sqrt6$ & & & & \\
$\bar{K}^0 \Sigma^-$ & $\sqrt2$ & & & & & & \\
${K}^- \Sigma^0$ & $-2$ & & & & & & \\
\end{tabular}
\end{ruledtabular}
\caption{Couplings of the anti-decuplet with the baryon octet and
pseudoscalar octet meson. Multiplying the universal coupling constant
$g$ is understood.}
\label{table2}
\end{table}

As an application, we compute the decay widths of the pure anti-decuplet
members.
Using the SU(3) Lagrangian, the total decay widths of the $N_{10}$,
$\Sigma_{10}$, and $\Xi_{10}$ are approximately $4 \times \Gamma_\Theta$,
$7 \times \Gamma_{\Theta}$, and $10 \times \Gamma_\Theta$, respectively,
which are expected to have some corrections if the decay channels into
vector meson plus baryons are opened.
As mentioned before, the $N_{10}$ cannot decay into $\Delta \pi$ because of
SU(3) flavor symmetry.
Furthermore, $U$-spin conservation does not allow its production from
$\gamma p$ reaction, while $\gamma n$ reaction can generate the
$N_{10}$.
Since the $N(1710)$ has large branching ratio of the decay into
$\Delta\pi$ and its coupling to $\gamma p$ is larger than that to
$\gamma n$ \cite{PDG02}, it cannot be a pure anti-decuplet state.
Therefore, there is no candidate for the $N_{10}$ among the known $J^P =
\frac12^+$ nucleon resonances \cite{PDG02}.
The $\Sigma(1770)$ with $J^P = \frac12^+$ may be a candidate
for the $\Sigma_{10}$.

However, if we consider the mixing with octet, the analyses become
more complex. The pentaquark octet can couple to meson octet and
baryon octet through the familiar $F$ and $D$ type interactions,
which introduces additional unknown coupling constants. It can
also couple to meson octet and baryon decuplet. So the mixing
debilitates the selection rules above. As far as the mass is
concerned, if we use the assumption that gives
(\ref{mass:numbers}), we found that the $N_q$ is much higher than
the $N(1440)$, but the $N_s$ is close to the $N(1710)$. For the
other states, we found that the $\Lambda(1600)$ with $\frac12^+$
can be a candidate for the $\Lambda_8$. As possible candidates for
$\Sigma_q$ and $\Sigma_s$, we found $\Sigma(1480)$ and
$\Sigma(2250)$ bumps in Particle Data Group \cite{PDG02} although
their quantum numbers are not fixed yet.

\medskip

Y.O. is grateful to the hospitality of Jefferson Lab., where this work
was completed.
This work was supported by Korea Research Foundation Grant
(KRF-2002-015-CP0074).

{\it Note added.\/} After completion of this work, we were aware
of a recent work of Diakonov and Petrov \cite{DP03}, which
discussed the mixing angles and similar mass relations in
pentaquark states.

\end{document}